\documentclass[aps,superscriptaddress,floatfix,onecolumn,preprint]{revtex4}

\usepackage{graphicx}

\usepackage{dcolumn}
\usepackage{bm}
\usepackage{setspace}
\usepackage{amsmath}

\newcommand{\beq}{\begin{eqnarray}}
\newcommand{\eeq}{\end{eqnarray}}

\usepackage{amssymb}

\begin{document}




\title{Studying the universality of field induced tunnel ionization times via high-order harmonic spectroscopy}


\author{H. Soifer$^{1}$, B.D. Bruner$^{1}$, M. Negro$^{2}$, M. Devetta$^{2}$, D. Faccial\`{a}$^{3}$, C. Vozzi$^{2}$, S.
de Silvestri$^{3}$, S. Stagira$^{3}$, and N. Dudovich}

\address{Department of Physics of Complex Systems, Weizmann
Institute of Science, Rehovot 76100, Israel\\
$^2$Institute for Photonics and Nanotechnologies, CNR, Milano, Italy\\
$^3$Department of Physics, Politecnico di Milano, Milano, Italy}

\date{\today}

\begin{abstract}
High-harmonics generation spectroscopy is a promising tool for resolving electron dynamics and structure in atomic and molecular systems. This scheme, commonly described by the strong field approximation, requires a deep insight into the basic mechanism that leads to the harmonics generation. Recently, we have demonstrated the ability to resolve the first stage of the process -- field induced tunnel ionization -- by adding a weak perturbation to the strong fundamental field. Here we generalize this approach and show that the assumptions behind the strong field approximation are valid over a wide range of tunnel ionization conditions.
Performing a systematic study -- modifying the fundamental wavelength, intensity and atomic system --  we observed a good agreement with quantum path analysis over a range of Keldysh parameters.
The generality of this scheme opens new perspectives in high harmonics spectroscopy, holding the potential of probing large, complex molecular systems.
\end{abstract}

%
%


\maketitle

\section{Introduction}\label{sec:intro}

Strong field light matter interactions open new horizons in ultrafast physics, in the generation of coherent XUV or x-ray radiation and energetic electrons. Understanding their basic mechanism is an essential step in establishing these new directions of research. In this framework an extensively studied phenomenon is high harmonic generation (HHG). In the strong field regime the mechanism that leads to the emission of high harmonics can be described in semiclassical terms. Near the peak of the optical cycle the Coulomb barrier is lowered by the strong laser field leading to tunnel ionization. Once the electron is free, it interacts with the laser field and accumulates high kinetic energy before it recollides with the parent ion and emits a high-energy photon \cite{Corkum93}. This semiclassical picture provides us with two important assumptions. It defines three \emph{separate} steps which can be treated independently. In addition, it provides a simple link between them: there is a direct mapping between the
time of ionization $t_0$ of the
electron, its time of return $t_r$ and the corresponding emitted photon energy $N\omega_0$ where $N$ is the harmonic order and $\omega_0$ is the fundamental frequency of the driving laser. These two important components of the semiclassical picture are at the heart of HHG spectroscopy. In this picture the tunneling serves as a 'pump', initiating a hole wavefunction, while the recollision serves as a 'probe', mapping the evolved wavefunction onto the HHG spectrum \cite{Ita04, Smirnov09, bucksbaum08, Bak06}.

Recently we have demonstrated the ability to decouple the two important steps in the process -- the ionization and recollision \cite{Shafir2012}. By adding a weak second harmonic field, orthogonally polarized with respect to the strong driving field, we induce a subcycle amplitude gate. Manipulating the two color delay shifts the gate within the optical cycle, probing a narrow range of electron trajectories according to their ionization time. A perturbative analysis enabled us to map each harmonic number to its time of ionization $t_0$ and time of recollision $t_r$. Our results, measured in helium atoms, demonstrated a deviation from the semi-classical picture, in agreement with quantum path analysis \cite{Lewen94}. Additionally, we have demonstrated the reconstruction of ionization times in both short and long trajectories in argon \cite{Soifer2013}. The extension of our approach to more complex electronic systems raises the following question: what are the limitations of the strong field assumptions? Specifically,
can we generalize our results to a wider range of fundamental parameters such as the laser frequency, intensity or atomic ionization potential?

The semiclassical picture provides a simple answer: there is a unique mapping between $t_0$, $t_r$ and $N\omega_0/U_p$ , where $U_p=\frac{E_0^2}{4\omega_0^2}$ is the pondermotive potential and $E_0$ is the field amplitude (atomic units are used throughout the paper). Such mapping provides a universal scaling law, in which the cutoff photon energy scales linearly with $U_p$. A more accurate picture is provided by the quantum mechanical description \cite{Lewen94}. Quantum mechanically, the induced dipole moment is dictated by all quantum trajectories that contribute to this process. These trajectories are coherently integrated with a relative phase known as the semi-classical action $S(\mathbf{p},t_0,t_r)$, given by:
\begin{align}
\label{action}
S(\mathbf{p},t_0,t_r)= \int_{t_0}^{t_r} dt \left(\frac{\left[\mathbf{p}-\mathbf{A}(t)\right]^2}{2}+I_p \right)
\end{align}
where $\mathbf{p}$ is the canonical momentum of the electron, $\mathbf{A}$ is the vector potential of the driving electric field and $I_p$ is the ionization potential of the target atom.

In the strong field regime the main contribution to the integral comes from the stationary points -- the points for which $\frac{\partial S}{\partial\mathbf{p}}=0$, $\frac{\partial S}{\partial t_r}=0$ and $\frac{\partial S}{\partial t_0}=0$. The stationary solutions $t_0^{st}$, $t_r^{st}$ and $\mathbf{p}^{st}$ provide us with a direct mapping between time and energy. These solutions differ from their semiclassical counterparts: the tunneling step, dictated by the ionization potential $I_p$ and the field's intensity $U_p$, couples the three main steps of the recollision process. As a direct consequence, an imaginary component appears in the stationary solutions -- $t_0^{st}$, $t_r^{st}$ and $\mathbf{p}^{st}$. Furthermore, the mapping between time and energy cannot be directly scaled since it is dictated by all parameters of the interaction.

\begin{figure}
\begin{center}
\includegraphics[width=0.45\textwidth]{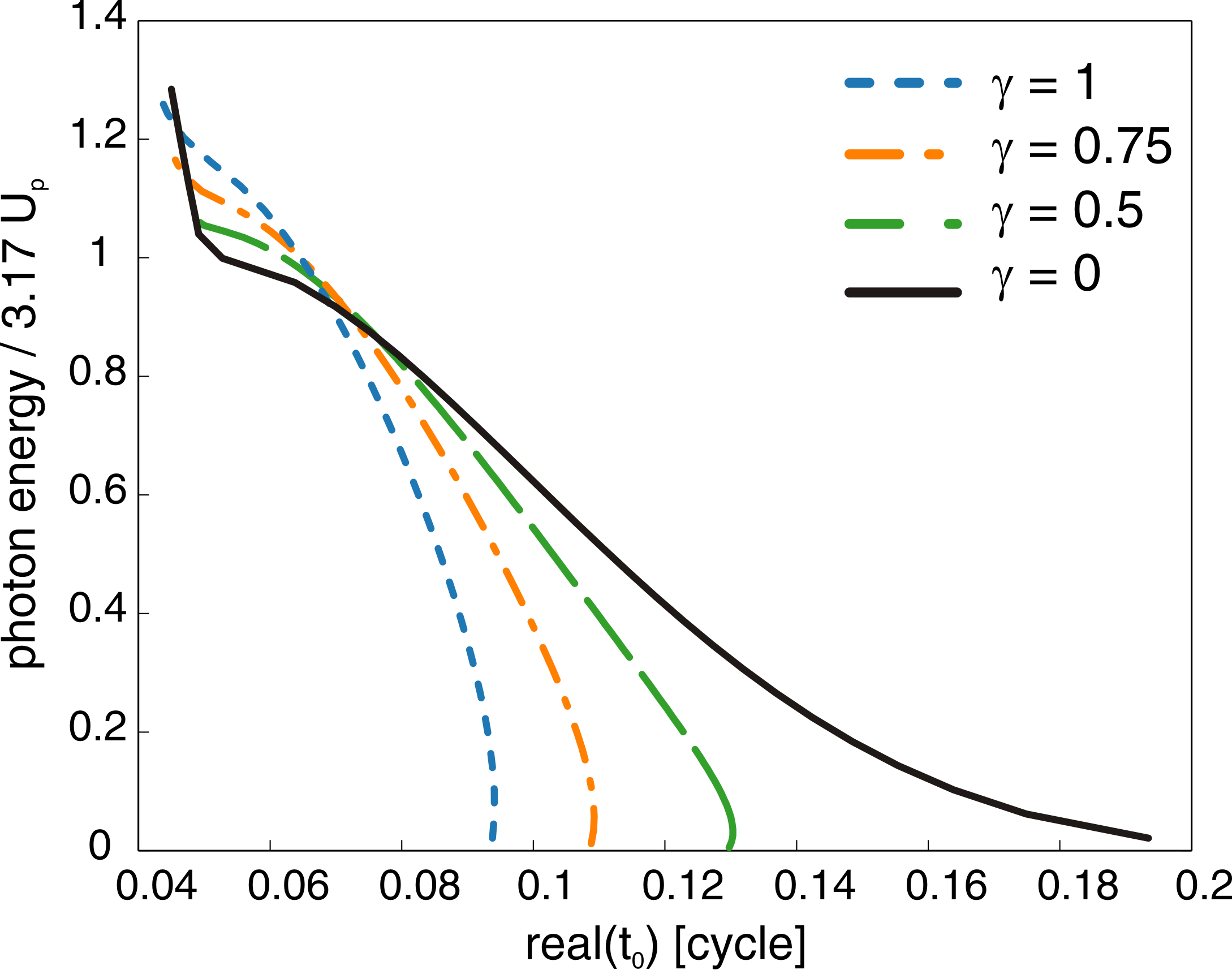}
\caption{Calculation of photon energy vs ionization time for various Keldysh numbers, namely $\gamma=0.5$ (green short-dashed curve), $\gamma=0.75$ (red dash-dotted curve) and $\gamma=1$ (blue dashed curve). The photon energy axis is normalized such that the classical cutoff is 1: $H_{norm}$=(photon energy - $I_p$)/3.17$U_p$. The real part of the ionization time is displayed in units of the optical cycle. With this normalization, the theoretical curves depend only on the Keldysh parameter, and not on $I_p$, $U_p$ and $\omega_0$ separately. The limit of $\gamma=0$ is shown as a black solid curve.}
\label{fig:1 times calc}
\end{center}
\end{figure}

We can characterize the interaction by the well known Keldysh parameter defined as $\gamma=\sqrt{\frac{I_p}{2U_p}}$ \cite{Keldysh}. This parametrization compares the two fundamental time scales in the process: the tunneling time $\tau=\frac{\sqrt{2I_p}}{E_0}$ and the fundamental period of the laser field $1/\omega_0$. Figure \ref{fig:1 times calc} describes the mapping between the normalized photon energy ($N\omega_0/3.17U_p$) and $Re\{t_0\}$ for a range of Keldysh parameters. In the limit of $\gamma\ll1$ ($I_p\ll U_p$) the stationary solution (solid black line)
approaches the classical one with the exception of the cutoff region, where the quantum nature of the process allows solutions which are not classically reproducible. Increasing $\gamma$ increases the difference between $Re\{t_0\}$ and the classical solution leading to a substantial narrowing of the ionization window.

In this paper we study the link between the ionization time and emitted photon energy for a range of Keldysh parameters between 0.64 and 1.23. Current laser technology enables us to achieve this range by scanning the fundamental wavelength \cite{Dimauro08}. In addition, we also modify the field intensity and vary the atomic systems, thus changing the ionization potential. To avoid contributions of multi-electron dynamics we focus the study to small, rare-gas, atomic systems. Our systematic study demonstrates the generality of the gating approach. Importantly, we confirm the validity of quantum path analysis in a regime were the basic assumptions that underly the strong field model (i.e. $\gamma\ll1$), reach their limit.

\section{methods}\label{sec:methods}
In the following we summarize the basic mechanism of the gating measurement. Consider a two color field in orthogonal configuration: $\mathbf{E}_{tot}(t)=E_0\left[\cos(\omega_0t)\hat{x}+\epsilon \cos(2\omega_0t+\phi)\hat{y}\right]$, where $E_0$ is the amplitude of the fundamental field, $\epsilon$ is the amplitude ratio between the two colors and $\phi$ is the two colors delay. In the limit of $\epsilon\ll1$ the second harmonic field acts as a weak perturbation inducing a small lateral shift of the recolliding electron. This shift reduces the recollision probability, suppressing the emitted signal. The lateral shift
accumulated between $t_0$ and $t_r$ can be classicaly described as \cite{Shafir2012}:
\begin{align}
\label{eq:displ}
y_L= \left(v_{0y}+\frac{\epsilon E_0}{2\omega_0}\sin(2\omega_0 t_0+\phi)\right)(t_r-t_0)+\frac{\epsilon E_0}{4\omega_0^2}\left(\cos(2\omega_0 t_r+\phi)-\cos(2\omega_0 t_0+\phi)\right).
\end{align}
Here $v_{0y}$ is the lateral electron velocity at the moment of ionization. The recollision condition which requires $y_L=0$ at $t=t_r$ defines the initial velocity $v_{0y}(t_0,t_r)$ that compensates for the field induced shift. The probability of an electron to tunnel with an initial velocity $v_{0y}(t_0,t_r)$ is centered around $v_{0y}=0$, having a width dictated by the tunneling process. Controlling the two colors delay, $\phi$, modulates the lateral shift and therefore $v_{0y}$. Its optimal value, $\phi_{max}$, for which the signal maximizes, is set at $v_{0y}(t_0,t_r)=0$, encoding the main parameters of the interaction: $t_0$ and $t_r$.

\begin{figure}
\begin{center}
\includegraphics[width=0.65\textwidth,clip]{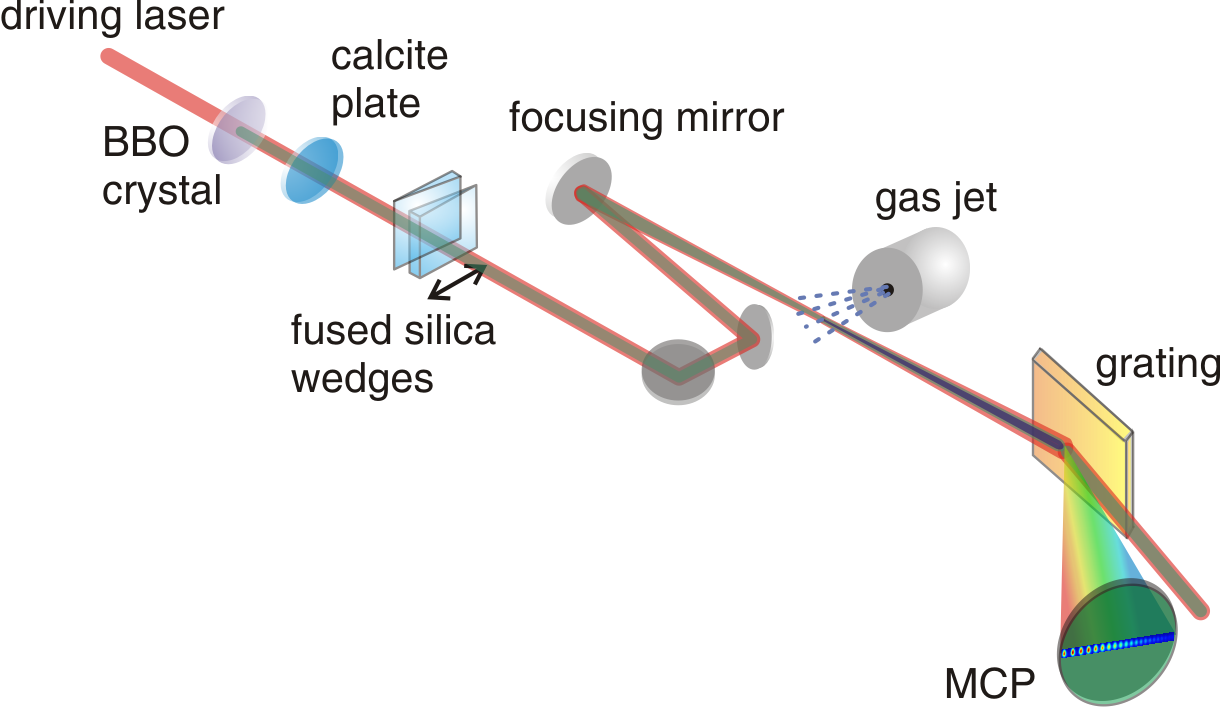}
\caption{General sketch of the experimental setup for two color HHG spectroscopy. BBO - $\beta$-Barium Borate. MCP - Micro Channel Plate}
\label{fig:2 experimental setup}
\end{center}
\end{figure}

Our systematic study was performed both in the 800 nm and the mid-IR regime.  Figure \ref{fig:2 experimental setup} shows a schematic description of the experimental setup.
We used a BBO crystal followed by a calcite window for generating the two color driving field, consisting of a strong HHG-generating field with perpendicular polarization relative to the weak second harmonic field. The relative phase of the two fields was controlled with a pair of glass wedges. High order harmonics were generated by focusing the two-color driving field on a pulsed gas jet. Harmonics were detected by means of an XUV spectrometer coupled to an MCP and a CCD detector. The 800 nm experiments were driven by 30 fs pulses generated by an amplified Ti:sapphire laser system.
The experiment driven by mid-IR  pulses (20 fs, tunable between 1300 and 1800 nm) were performed exploiting an optical parametric amplifier (OPA) pumped by an amplified Ti:sapphire laser system (60 fs, 20 mJ, 800 nm) \cite{Vozzi2007}.

\begin{figure}
\begin{center}
\includegraphics[width=0.65\textwidth,clip]{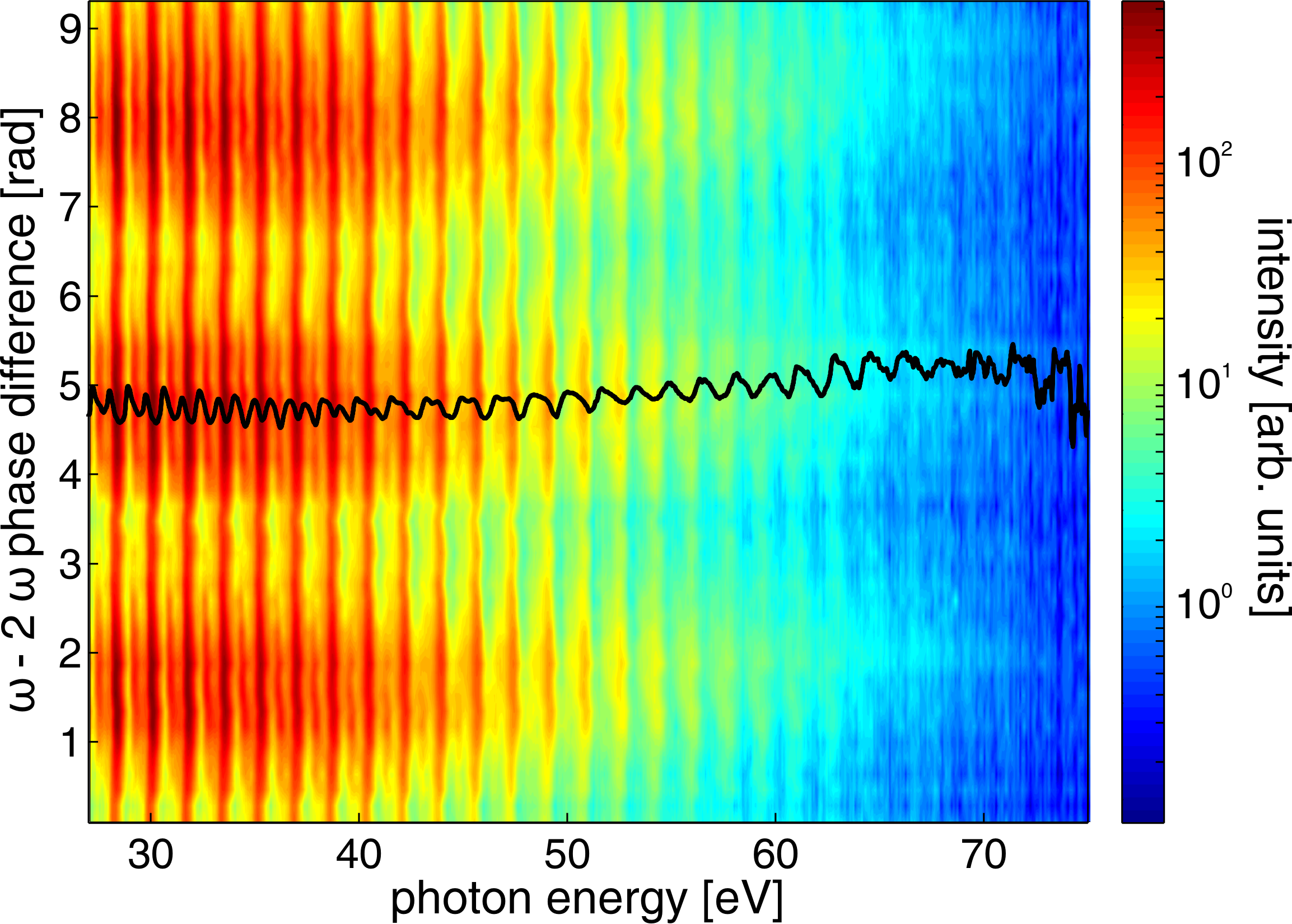}
\caption{Sequence of harmonic spectra acquired in krypton at $\lambda$= 1450 nm as a function of photon energy and two-color phase difference (log scale). The black solid line represents the two-color delay for which the signal is maximized ($\phi_{max}$).}
\label{fig:3 2D}
\end{center}
\end{figure}

\section{results}\label{sec:results}
Figure \ref{fig:3 2D} shows an example of the experimental results: normalized harmonic spectra driven in krypton by 1450-nm pulses are shown as a function of the two colors delay, $\phi$. The solid curve follows $\phi_{max}$ as a function of the harmonic order. As can be clearly observed $\phi_{max}$ changes with the harmonic order reflecting the subcycle nature of the gating measurement. Indeed, as in previous studies, in the IR regime a good agreement with quantum path analysis is obtained. Analogous experimental results were obtained with 800-nm driving pulses.

\begin{figure}
\begin{center}
\includegraphics[width=0.45\textwidth]{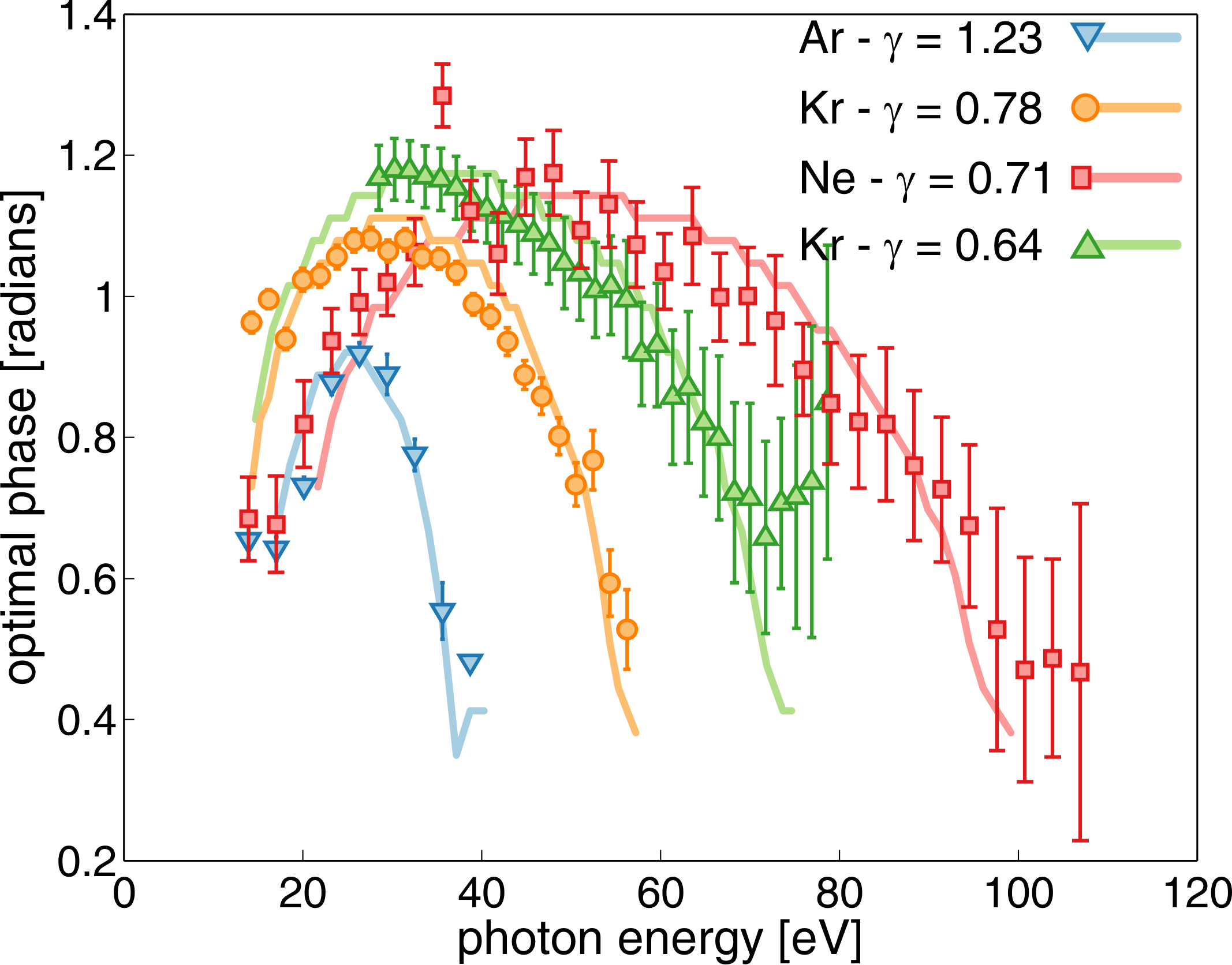}
\caption{Optimal phase $\phi_{max}$ vs photon energy. Experimental results are represented by scatter graphs and theoretical calculation are represented by curves. The experimental points were shifted in the vertical axis to best fit the theoretical curve, since the two-color delay $\phi$ is only known up to a constant shift. The errorbars represent the uncertainty in fitting the modulation in intensity of each harmonic order to a sinusoidal function. The four data sets correspond to the following parameters: (1) blue downward triangles - Ar, $\lambda=800$ nm,  $I_p=15.8$ eV, $U_p=5.2$ eV, $\gamma=1.23$. (2) orange circles - Kr, $\lambda=1300$ nm,  $I_p=14.0$ eV, $U_p=11.5$ eV, $\gamma=0.78$. (3) red squares - Ne, $\lambda=800$ nm,  $I_p=21.6$ eV, $U_p=21.1$ eV, $\gamma=0.71$. (4) green upward triangles - Kr, $\lambda=1434$ nm,  $I_p=14.0$ eV, $U_p=16.9$ eV, $\gamma=0.64$.}
\label{fig:4 phase energy}
\end{center}
\end{figure}

We performed a systematic study of the gating experiment over a range of Keldysh parameters. Figure \ref{fig:4 phase energy} summarizes the results, displaying the two-color optimal delay $\phi_{max}$ vs photon energy for four values of Keldysh parameter. These data were obtained from three different atoms, and with three wavelengths, as follows: (1) Argon, 800nm, $\gamma=1.23$. (2) Krypton, ~1300nm, $\gamma=0.78$ (3) Neon, 800nm, $\gamma=0.71$, and (4) Krypton, ~1430nm, $\gamma=0.64$.
The two-color phase is only known experimentally up to a global shift. Therefore, the four curves were shifted in the vertical direction to best fit the theoretical prediction. The theoretical curves (solid lines) were calculated by solving the gating equation (eq. \ref{eq:displ}) for $\phi$, setting $y_L=0$, $v_{0y}=0$, based on the ionization times obtained from the quantum path analysis \cite{Lewen94,Zhao2013}.
Since the peak intensity of the laser was not accurately known, $U_p$ was an additional fitting parameter in calculating the theoretical curves. Our study demonstrates a good agreement with quantum path analysis within this range of Keldysh parameters.

\begin{figure}
\begin{minipage}[b]{0.45\linewidth}
\centering
\includegraphics[width=\textwidth]{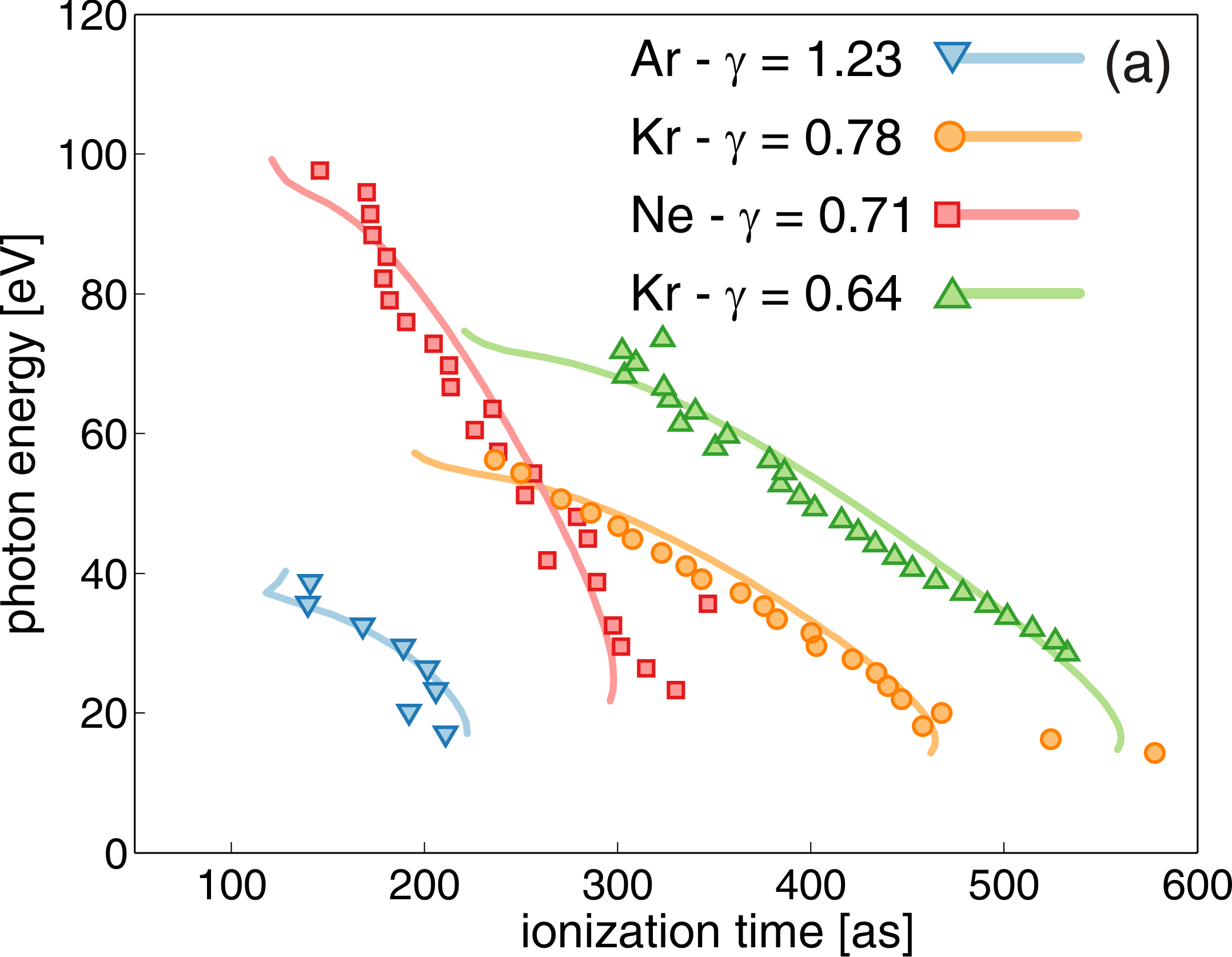}
\label{fig:figure5a}
\end{minipage}
\hspace{0.5cm}
\begin{minipage}[b]{0.458\linewidth}
\centering
\includegraphics[width=\textwidth]{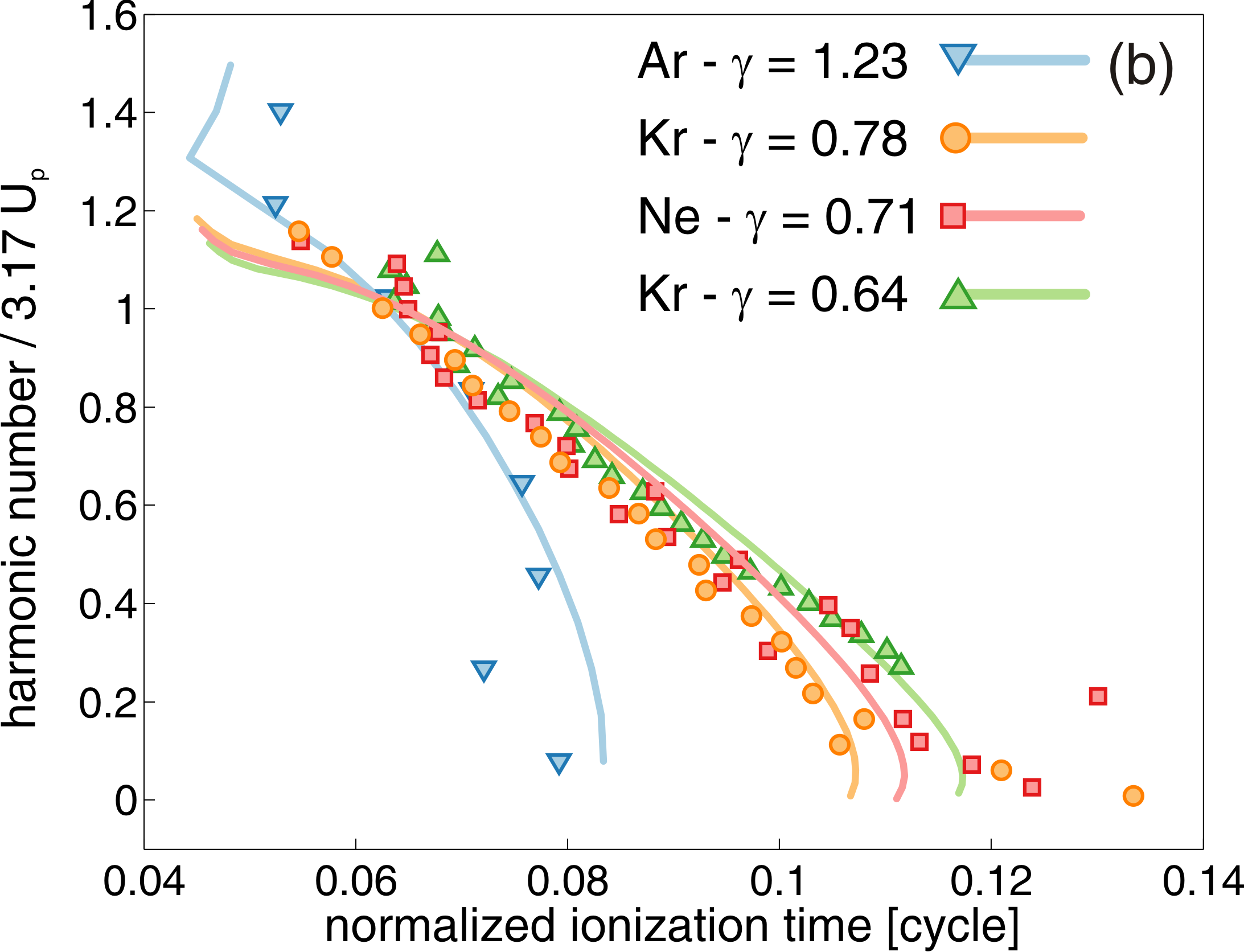}
\label{fig:figure5b}
\end{minipage}
\caption{Reconstructed ionization times extracted from the experimental points ($\phi_{max}$) in figure \ref{fig:4 phase energy}. The solid curves represent the theoretical times $Re\{t_0^{st}\}$. In panel (a) it can be observed that the curves are spread in energy and time due to the different experimental conditions. Panel (b) shows the same data in normalized units: photon energy $H_{norm}=($photon energy $ - I_p)/3.17U_p$ and ionization time in units of the optical cycle. The normalized ionization times for measurements with close keldysh numbers ($\gamma=0.64-0.78$) are very similar, whereas the measurement with significantly larger keldysh number ($\gamma=1.23$) indeed has a higher cutoff and a narrower ionization window.}
\label{fig: 5}
\end{figure}

In the next stage, we map the information from the spectral domain to the temporal domain and reconstruct the ionization times. The reconstruction procedure is described in detail in \cite{Soifer2013}. Specifically, our experiment resolves $\phi_{max}$ as a function of the harmonic number for each value of Keldysh parameter. Relying on the perturbative approach we calculate the lateral gate and extract the theoretical $\phi_{max}(t_r,t_0)$. Comparing $\phi_{max}(N\omega_0)$ to $\phi_{max}(t_r,t_0)$ and relying on the theoretical values of $t_r$ (which have been independently measured before \cite{atto_Misha,Doumy2009}) we extract $Re\{t_0\}$. Figure \ref{fig: 5} (a) shows the results of this reconstruction, as well as the calculated ionization times $Re\{t_0^{st}\}$. As expected, the reconstructed curves are spread both in energy and in time due to the large range of $U_p$, $I_p$ and wavelength.

In order to compare these different experiments, we show in Fig. \ref{fig: 5} (b) the same data as a function of normalized time and energy axes. The ionization time is normalized according to the duration of the optical cycle, such that measurements performed with different driving wavelengths are on the same scale. In addition, the photon energy scale is normalized for each measurement in the following way: $H_{norm}=($photon energy $ - I_p)/3.17U_p$ (such that the classical cutoff is 1). With this normalization, the curves of $t_0^{st}$ depend only on the Keldysh parameter, whereas the classical times are independent of $I_p$, $U_p$ or wavelength, as shown theoretically in fig. \ref{fig:1 times calc}. All experimental curves deviate significantly from the classical curve. It can be observed that the normalized ionization times for measurements with close Keldysh numbers ($\gamma=0.64-0.78$) are very similar, whereas the measurement with significantly larger Keldysh parameter ($\gamma=1.23$) has a narrower ionization window.

These results display a striking universality. Although obtained with a large range of experimental parameters - varying $I_p$, peak intensity and wavelength - they can all be described by a curve which depends only on the Keldysh parameter. Importantly, we obtained a very good fit even for a measurement with $\gamma=1.23$. This observation demonstrates the validity of quantum path analysis of HHG in a regime where the adiabatic assumption of $\gamma\ll1$ fails \cite{Lai2013}.

Previous experiments have measured the attochirp in HHG generated by mid-IR lasers \cite{Doumy2009}. In this work we report the first measurement of ionization times at wavelengths longer than 800 nm. Combined with our previous results\cite{Shafir2012, Soifer2013}, we confirm the validity of the strong field approximation and the single-electron picture for a variety of atomic systems and a wide range of laser parameters. This is an important step for extending HHG spectroscopy to mid-IR wavelengths, a regime which holds great promise \cite{Vozzi2012, Dichiara2013}.

Generalizing the two-color gating scheme to longer wavelengths enables extending our scheme to resolve tunneling dynamics in more complex molecular systems. In these systems structural and dynamical effect are often coupled \cite{Worner10}, adding a major obstacle to HHG molecular spectroscopy \cite{Smirnov09, Vozzi2011}. In addition, the tunneling process may involve nontrivial multielectron dynamics. Extending the dimensionality of the measurement by combining two-color schemes with tuneable mid-IR sources will enable us to disentangle the main degrees of freedoms,  resolving attosecond scale processes, so far hidden in many HHG experiments.

\section*{Acknowledgements}
The research leading to these results has received funding from LASERLAB-EUROPE (grant agreement n° 284464, EC Seventh Framework Programme),from ERC Starting Research Grant UDYNI (grant agreement n° 307964, EC Seventh Framework Programme)
and from the Italian Ministry of Research and Education (ELI project - ESFRI Roadmap). N.D. acknowledges the Minerva Foundation, the Israeli Science Foundation, the German-Israeli Foundation, the European Research Council foundation, the Crown Center of Photonics and the European Research Council for financial support. H. S. is supported by the Adams Fellowship
Program of the Israel Academy of Sciences and Humanities.

\bibliographystyle{naturemag}

\bibliography{bibliotheque2}

\end{document}